\documentclass[11pt,english]{article}
\usepackage{mathptmx}

\usepackage[T1]{fontenc}
\usepackage[latin9]{inputenc}
\usepackage[letterpaper]{geometry}
\usepackage{amsmath}
\usepackage{graphicx}
\usepackage{esint}
\usepackage[numbers,comma,sort&compress]{natbib}

\makeatletter

\providecommand{\tabularnewline}{\\}

\usepackage{amsfonts}
\usepackage{dsfont}

\usepackage{times}
\usepackage{calc}

\newcommand{\beq}{\begin{equation}}
\newcommand{\eeq}{\end{equation}}

\usepackage{babel}

\makeatother

\usepackage{babel}
\begin{document}
\title{Physics of the Edwards-Anderson Spin Glass in Dimensions $d=3,\ldots,8$
from Heuristic Ground State Optimization}
\author{Stefan Boettcher\thanks{http://www.physics.emory.edu/faculty/boettcher},
Physics Department, Emory University, Atlanta, Georgia 30322, USA }
\maketitle
\begin{abstract}
We present a collection of simulations of the Edwards-Anderson lattice
spin glass at $T=0$ to elucidate the nature of low-energy excitations
over a range of dimensions that reach from physically realizable systems
to the mean-field limit. Using heuristic methods, we sample ground
states of instances to determine their energies while eliciting excitations
through manipulating boundary conditions. We exploit the universality
of the phase diagram of bond-diluted lattices to make such a study
in higher dimensions computationally feasible. As a result, we obtain
a verity of accurate exponents for domain wall stiffness and finite-size
corrections that allow us to examine their dimensional behavior and
their connection with predictions from mean-field theory. We also
provide an experimentally testable prediction for the thermal-to-percolative
crossover exponent in dilute lattices Ising spin glasses. 
\end{abstract}

\section{Introduction\label{sec:Introduction}}

Imagining physical systems in non-integer dimensions, such as through
the $\epsilon$-expansion~\citep{Wilson72} or \textit{dimensional
regulation}~\citep{tHooft72}, to name but two, has provided many
important results for the understanding of the physics in realistic
dimensions. For example, the goal of the $\epsilon$-expansion is
to establish a connection between the (technically, infinite-dimensional)
mean-field solution of a field theory and its real-space behavior.
For a disordered system such as a spin glass~\citep{F+H,MPV,Stein13,RSB40},
this playbook has proved rather difficult to follow theoretically
\citep{dedominicis:98,Moore18,Moore21}. In contrast, we endeavor
to explore the transition between the often well-known mean-field
properties and their modifications in real space using numerical means,
free of any theoretical preconceptions. In this task, on top of the
computational extensive disorder averages, the complexity of spin
glasses reveals itself through increasingly slower convergence in
thermal simulations, the deeper one pushes into the glassy regime.
Going all the way to $T=0$, then, makes thermal explorations impossible
and renders the problem of finding the ground states NP-hard in general
\citep{Barahona82}, in fact. However, simulations at $T=0$ also
avail us considerable conceptual clarity and an entirely new suit
of techniques, albeit for just a few, yet important, observables.
Some equilibrium properties of spin glasses below $T_{c}$ can be
obtained from merely determining ground state energies, such as domain
wall stiffness, finite-size corrections, and thermal-percolative crossover
exponents. To keep systematic errors low while also creating enough
statistics for the disorder average, we need to employ fast but ultimately
inexact heuristic methods to overcome NP-hardness. To reach a sensible
scaling regime in system sizes $N$, especially in higher dimensions,
also requires clever exploitation of the phase diagram of a spin glass.
Here, we discuss together the results obtained from large-scale simulations
conducted over several years and spread over a number of articles
\citep{Boettcher04b,Boettcher04c,Boettcher05d,Boettcher07a,BoFa11}.

To be specific, we simulate the Ising spin glass model due to Edwards
and Anderson (EA) with the Hamiltonian \citep{Edwards75} 
\begin{eqnarray}
H & = & -\sum_{<i,j>}\,J_{i,j}\,\sigma_{i}\,\sigma_{j}.\label{eq:Heq}
\end{eqnarray}
The dynamic variables are binary (Ising) spins $\sigma_{i}=\pm1$
placed on a hyper-cubic lattice in integer $d$ dimension with couplings
between nearest neighbors $<i,j>$ via random bonds $J_{ij}$ drawn
from some distribution ${\cal P}(J)$ of zero mean and unit variance.
The lattices are periodic with base length $L$ in all directions,
i.e., each such instance has $N=L^{d}$ spins. To relate real-world
behavior in $d=3$ (which is explored experimentally and theoretically
in other articles in this collection) with mean-field behavior, which
manifests itself above the upper critical dimension $d_{u}=6$ \citep{F+H},
we find ground states of EA on lattices in $d=3,\ldots,8$. In each
$d$, we need to simulate instances over a wide range of $L$ to be
able to extrapolate our results to the thermodynamic limit ($L\to\infty$).
At each size $L$, we further need to measure a large number of instances
with independently drawn random bonds for the disorder average inherent
to obtain observables in spin glasses. Each instance entails approximating
its ground state, which is an NP-hard combinatorial problem.

\begin{figure}
\begin{minipage}[b][1\totalheight][s]{0.48\textwidth}%
\caption{\label{fig:PhaseDia}{\small{}{} Phase diagram for bond-diluted spin
glasses ($d>d_{l}$). The entire spin-glass phase (SG) for $T<T_{c}$
and $p>p_{c}$ has a universal positive domain-wall exponent, $y>0$.
In our measurements, we therefore utilize an interval of bond densities
at $T=0$ (red arrow) where $p$ is sufficiently above the scaling
window near $p_{c}$ (at finite system size) but small enough to asymptotically
reach significant system sizes $L$. At $p=p_{c}$ and $T=T_{c}=0$,
we define the domain-wall exponent for a spin glass on the percolating
cluster as $y=y_{P}(<0)$. It allows to extract the thermal-percolative
crossover exponent $\phi$ that describes the behavior along the boundary
$T_{c}(p)\sim(p-p_{c})^{\phi}$ for $p\searrow p_{c}$ (green arrow).
In the paramagnetic phase (PM) for $p<p_{c}$ or $T>T_{c}$, defect
energies due to domain walls decay exponentially.}}
\end{minipage}\hfill{}%
\begin{minipage}[b][1\totalheight][s]{0.48\textwidth}%
\includegraphics[bb=60bp 100bp 500bp 470bp,clip,width=1.05\textwidth,viewport=50bp 120bp 500bp 470bp]{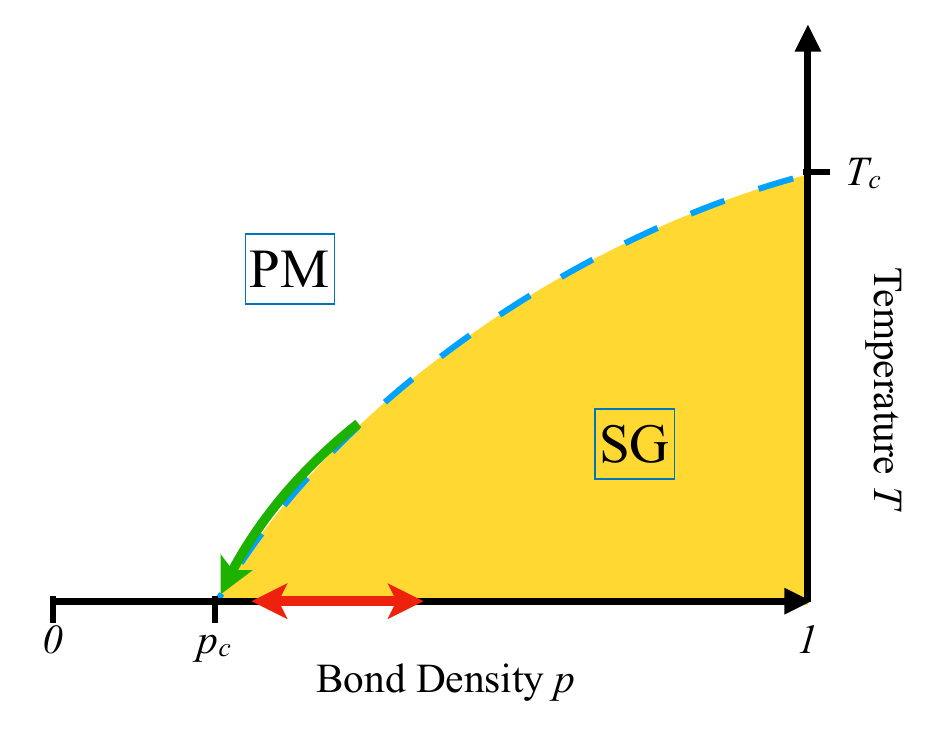}\hfill{}%
\end{minipage}\hfill{} 
\end{figure}

To sample ground state energies at high through-put and with minimal
systematic error, heuristics can only be relied on for systems with
not too much more than $N\approx1000$ spins coupled together. This
would appear to limit the ``dynamic range'' in size up to around
$L=10$ in $d=3$, but limited to $L=6$ in $d=4$ and even to $L<3$
in $d=7$, definitely insufficient to extract any $L\to\infty$ limit!
However, the phase diagram for a bond-diluted EA system (with $d\geq3$
such that $T_{c}>0$) in Fig. \ref{fig:PhaseDia} suggests that universal
scaling behavior extends across the entire spin-glass phase (SG) down
to the scaling window near the bond-percolation threshold $p_{c}$
for low enough $T$, i.e., most definitely for $T=0$. Thus, our strategy
is to find ground states for EA instances at bond-density $p$ with
sufficient dynamic range in $L$ for $p>p_{c}$ just above that scaling
window to be within SG, using \emph{exact} reduction methods \citep{Boettcher04c,Boettcher08a}
(see Appendix A) to remove a large number of spins followed by heuristic
optimization of remainder systems with $N_{r}\leq1000$ \citep{Boettcher01a,Dagstuhl04}
(see Appendix B). These reduction methods recursively trace out all
spins with fewer than four connected neighbors, at least, and are
particularly effective near $p_{c}$, since each spin in EA has at
most $2d$ potential neighbors while $p_{c}\sim1/(2d)$ in large $d$
such that for $p$ just above $p_{c}$ lattices remain sparse, each
spin being connected to barely more than one other spin, on average,
albeit with large variations. E.g., in $d=8$ for $p=0.0735>p_{c}\approx0.068$
and $L=6$, an EA system with $N=6^{8}\approx1.7\times10^{6}$ spins
typically reduces to a remainder graph with $\left\langle N_{r}\right\rangle \approx1000$
spins, each connected to 5.3 neighbors, on average, to be optimized
heuristically.

\section{Domain Wall Stiffness Exponents\label{sec:Domain-Wall-Stiffness}}

A quantity of fundamental importance for the modeling of amorphous
magnetic materials through spin glasses~\citep{F+H,Southern77,McMillan84,Fisher86,bray:86}
is the domain-wall or ``stiffness'' exponent $y$, often also labeled
$\theta$. As Hook's law describes the response to increasing elastic
energy imparted to a system for increasing displacement $L$ from
its equilibrium position, the stiffness of a spin configuration describes
the typical rise in magnetic energy $\Delta E$ due to an induced
defect-interface of a domain of size $L$. But unlike uniform systems
with a convex potential energy function over its configuration space
(say, a parabola for the single degree of freedom in Hook's law, or
a high-dimensional funnel for an Ising ferromagnet), an amorphous
many-body system exhibits a function more reminiscent of a high-dimensional
mountain landscape. Any defect-induced displacement of size $L$ in
such a complicated energy landscape may move a system through numerous
undulations in energy $\Delta E$. Averaging over many incarnations
of such a system results in a typical energy scale 
\begin{eqnarray}
\sigma(\Delta E) & \sim & L^{y}\qquad(L\to\infty)\label{eq:yscaling}
\end{eqnarray}
for the standard deviations of the domain wall energy $\Delta E$.

The importance of this exponent for small excitations in disordered
spin systems has been discussed in many contexts~\citep{Fisher86,Krzakala00,Palassini00,Palassini03,Bouchaud03,Aspelmeier03}.
Spin systems with $y>0$ provide resistance (``stiffness'') against
the spontaneous formation of defects at sufficiently low temperatures
$T$; an indication that a phase transition $T_{c}>0$ to an ordered
state exists. For instance, in an Ising ferromagnet, the energy $\Delta E$
is always proportional to the size of the interface, i.~e. $y=d-1$,
consistent with the fact that $T_{c}>0$ only when $d>1$. For $y<0$,
the state of a system is unstable with respect to defects, and spontaneous
fluctuations may proliferate, preventing any ordered state. Thus,
determining the exact ``lower critical dimension'' $d_{l}$, where
$y\vert_{d=d_{l}}=0$, is of singular importance, and understanding
the mechanism leading to $d_{l}$, however un-natural its value, provides
clues to the origin of order \citep{Bray84,Franz94,Hartmann01,Boettcher05d,Guchhait14,Maiorano18}.

Instead of waiting for a thermal fluctuation to spontaneously induce
a domain-wall, it is expedient to directly impose domains of size
$L$ through reversed boundary conditions on the system and measure
the energy needed to determine $y$. To wit, in a system with periodic
boundary conditions, we first obtain its ground state $E_{0}$ unaltered
and obtain it again as $E_{0}^{\prime}$ after reversing the signs
on all bonds within a $(d-1)$-dimensional hyperplane, resulting in
a complex domain of spins of size $\sim L$ that are reversed between
both ground states such that $\Delta E=E_{0}-E_{0}^{\prime}$ is the
energy due to the interface of that domain. Since $\Delta E$ is equally
likely to be positive or negative, it is its deviation, $\sigma(\Delta E)$,
which sets the energy scale in Eq. (\ref{eq:yscaling}). Note that
this problem puts an even higher demand on the ground state heuristic
than described in the introduction. Here, the domain-wall energy $\Delta E$
is a minute, sub-extensive difference between two almost identical,
extensive energies, $E_{0}$ and $E_{0}^{\prime}$, each of which
is NP-hard to find. Thus, any systematic error would escalate rapidly
with $N_{r}$, the size of the remainder graph.

\begin{figure*}
\hfill{}\includegraphics[bb=0bp 200bp 800bp 618bp,clip,width=0.97\textwidth,viewport=0bp 200bp 800bp 618bp]{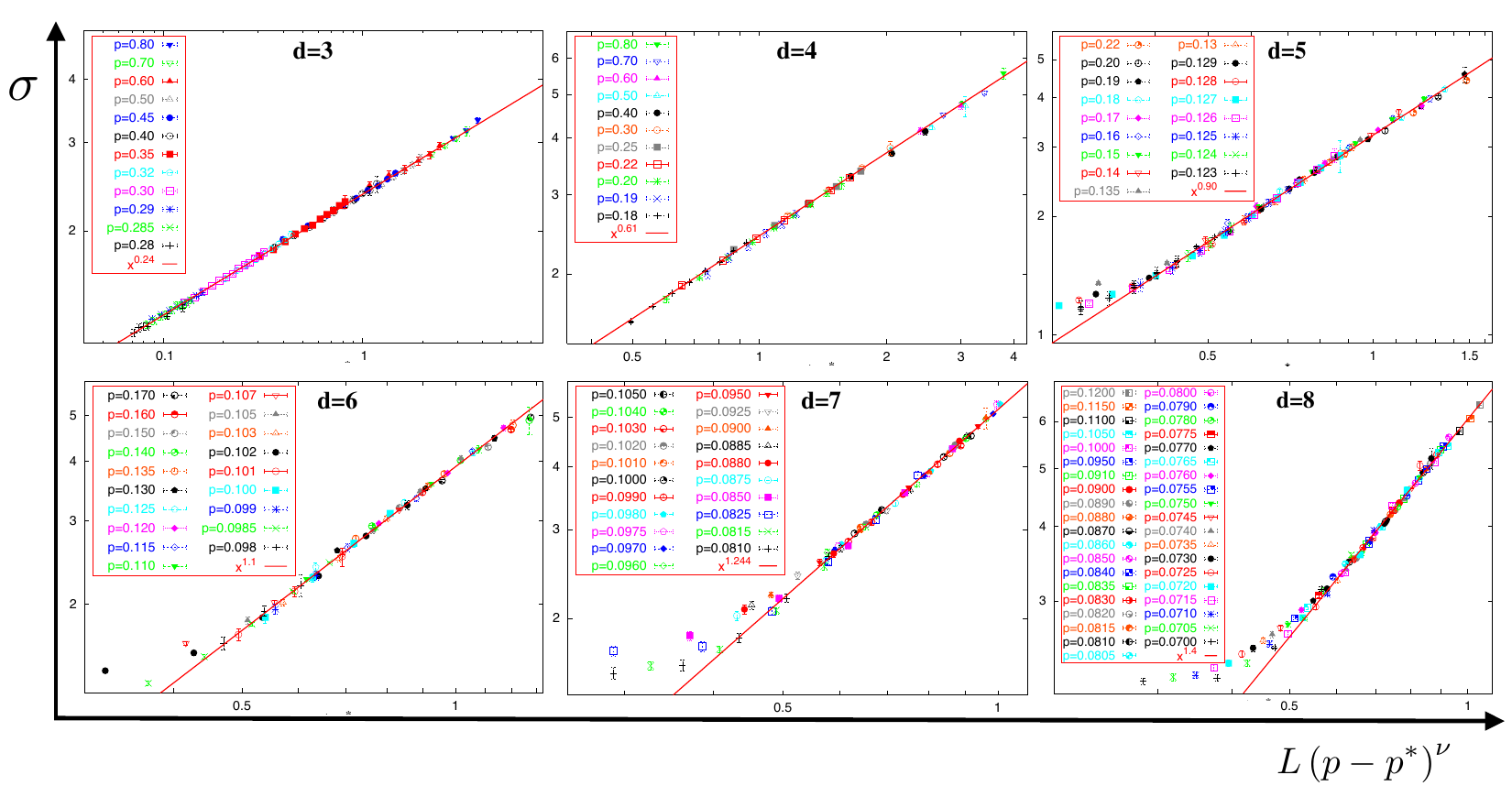}\hfill{}

\caption{\label{fig:StiffnessScaling} {\small{}{}Data collapse for the domain
wall scaling simulations of bond-diluted EA in $d=3,\ldots,8$ of
$\sigma(L,p)\sim{\cal L}^{y}$ in the scaling variable ${\cal L}=L\left(p-p^{*}\right)^{\nu}$.
For each $d$, data sets are created over a range in $p$ as listed
in the respective legend, up to a size $L$ such that remainder graphs
are typically $<\left\langle N_{r}\right\rangle \approx10^{3}$. The
original data and the fitting parameters are listed in Refs. \citep{Boettcher04b,Boettcher04c}.
The obtained domain wall scaling exponents $y_{d}$ are listed in
Tab. \ref{tab:Stiffness-exponents}. Note that for $d\protect\leq4$,
transient data for smaller ${\cal L}$ has been omitted for clarity.}}
\end{figure*}

As shown in Fig. \ref{fig:StiffnessScaling}, using bond-diluted lattices
for EA, in contrast, not only affords us a larger dynamic range in
$L$, but also allows for an extended scaling regime due to the additional
parameter of $p$ ranging over an entire interval. Instead of one
set of data for increasing $L$ at a fixed $p$ (typically, $p=1$
\citep{Hartmann2000}) leading to the scaling in Eq. (\ref{eq:yscaling}),
we can scale multiple independent sets for such a range of $p$ into
a collective scaling variable, ${\cal L}=L\left(p-p^{*}\right)^{\nu}$,
that collapses the data according to $\sigma(L,p)\sim{\cal L}^{y}$.
While the extension to an interval in $p$ makes simulations more
laborious, it typically yields an extra order of magnitude in scaling
compared to the prohibitive effort of confronting the NP-hard problem
of reaching large $L$ at fixed $p$ alone. For instance, in $d=3$
at $p=1$, attainable sizes span $3\leq L\leq12$, at best, while
we obtain a perfect data collapse for about $0.07\leq{\cal L}\leq3$
for $0.28\leq p\leq0.8$. (Note that while $p^{*}\approx p_{c}$ and
$\nu$ has some relation to the correlation-length exponent in percolation,
see below, it is necessary to allow these to be a free parameter for
the bimodal bonds used in these simulations, as was argued in Ref.
\citep{Boettcher04c}.) The fitted values for $y$ for each $d$,
as obtained from Fig. \ref{fig:StiffnessScaling}, are listed in Tab.
\ref{tab:Stiffness-exponents}.

These values for $y$ are listed in Tab. \ref{tab:Stiffness-exponents}
and plotted in Fig. \ref{fig:1overd} as $1-\frac{y}{d}$. That quantity
has been obtained in the mean-field case by Parisi and Rizzo \citep{Parisi08},
yielding $1-\frac{y}{d}=\frac{5}{6}$ above the upper critical dimension,
$d\geq d_{u}=6$. That value is clearly consistent with our high-dimensional
data, providing a rare direct comparison between the mean field theory
(RSB) and real-world spin glasses. As Fig.~\ref{fig:1overd} further
shows, the exponent varies continuously with dimension $d$ and allows
for a simple cubic fit of the numerical data between $2\leq d\leq6$,
weighted by the statistical errors \citep{Boettcher05d}. The fit
\textit{independently} reproduces the exactly-known result \emph{outside}
the fitted domain at $d=1$, $y=-1$, to less than $0.8\%$ (not shown
here). The fit has a zero at $d_{l}\approx2.498$ and yields $y\approx0.001$
at $d=\frac{5}{2}$; strong evidence that $d_{l}=5/2$, which has
been suggested also by theory \citep{Franz94,Maiorano18} and is consistent
with experiment \citep{Guchhait14}.

In the following, we will consider some other uses of the domain-wall
excitations.

\begin{table}[b!]
\caption{\label{tab:Stiffness-exponents}{\small{}{} Stiffness exponents for
Edwards-Anderson spin glasses \citep{Boettcher04b,Boettcher04c} for
dimensions $d=2,\ldots,8$ obtained numerically from domain-wall excitations
of ground states, as in Fig. \ref{fig:StiffnessScaling}. The next
column contains the measured values for finite-size corrections, denoted
as $\omega$, from the fit of the data in Fig.~\ref{fig:FSC}. The
stiffness exponents $y_{P}$ obtained in Ref. \citep{Boettcher07a}
refer to EA at the bond-percolation threshold $p_{c}$, with values
of $p_{c}$ taken from Ref.~\citep{Lorenz98} for $d=3$ and Ref.~\citep{Grassberger03}
for $d\protect\geq4$. The correlation-length exponents $\nu$ for
percolation are from Ref.~\citep{Deng05} in $d=3$ and from Ref.~\citep{Hughes96}
for $d\protect\geq4$, where $\nu=1/2$ is exact above the upper critical
dimension, $d\protect\geq6$.}}

\vspace{0.2cm}
 \hfill{}%
\begin{tabular}{|r@{\extracolsep{0pt}.}l|r@{\extracolsep{0pt}}l|r@{\extracolsep{0pt}.}l|r@{\extracolsep{0pt}}l|r@{\extracolsep{0pt}}l|r@{\extracolsep{0pt}.}l|r@{\extracolsep{0pt}}l|r@{\extracolsep{0pt}.}l|r@{\extracolsep{0pt}.}l}
\hline 
\multicolumn{2}{|c||}{$d$} & \multicolumn{2}{c|}{$y$} & \multicolumn{2}{c||}{$1-y/d$} & \multicolumn{2}{c||}{$\omega$} & \multicolumn{2}{c|}{$y_{P}$} & \multicolumn{2}{c||}{$1-y_{P}/d$} & \multicolumn{2}{c|}{$p_{c}$} & \multicolumn{2}{c|}{$\nu$} & \multicolumn{2}{c|}{$\phi=-\nu y_{P}$}\tabularnewline
\hline 
\hline 
\multicolumn{2}{|c||}{2} & \multicolumn{2}{c|}{-0} & 1&141(1)  & \multicolumn{2}{c||}{} & \multicolumn{2}{c|}{-0} & 1&497(2)  & \multicolumn{2}{c|}{$\frac{1}{2}$} & \multicolumn{2}{c|}{$\frac{4}{3}$} & 1&323(4)\tabularnewline
\hline 
\multicolumn{2}{|c||}{3} & \multicolumn{2}{c|}{0} & 0&920(4)  & \multicolumn{2}{c|}{0} & \multicolumn{2}{c|}{-1} & 1&429(3)  & \multicolumn{2}{c|}{0} & 0&87436(46)  & 1&127(5)\tabularnewline
\hline 
\multicolumn{2}{|c||}{4} & \multicolumn{2}{c|}{0} & 0&847(3)  & \multicolumn{2}{c|}{0} & \multicolumn{2}{c|}{-1} & 1&393(2)  & \multicolumn{2}{c|}{0} & 0&70(3)  & 1&1(1)\tabularnewline
\hline 
\multicolumn{2}{|c||}{5} & \multicolumn{2}{c|}{0} & 0&824(10)  & \multicolumn{2}{c|}{0} & \multicolumn{2}{c|}{-1} & 1&37(1)  & \multicolumn{2}{c|}{0} & 0&571(3)  & 1&05(2)\tabularnewline
\hline 
\multicolumn{2}{|c||}{6} & \multicolumn{2}{c|}{1} & 0&82(2)  & \multicolumn{2}{c|}{0} & \multicolumn{2}{c|}{-2} & 1&34(1)  & \multicolumn{2}{c|}{0} & \multicolumn{2}{c|}{$\frac{1}{2}$} & 1&00(2)\tabularnewline
\hline 
\multicolumn{2}{|c||}{7} & \multicolumn{2}{c|}{1} & 0&823(7)  & \multicolumn{2}{c|}{0} & \multicolumn{2}{c|}{-2} & 1&33(1)  & \multicolumn{2}{c|}{0} & \multicolumn{2}{c|}{$\frac{1}{2}$} & 1&14(3)\tabularnewline
\hline 
\multicolumn{2}{|c||}{8} & \multicolumn{2}{c|}{1} & 0&85(2)  & \multicolumn{2}{c||}{} & \multicolumn{2}{c|}{} & \multicolumn{2}{c||}{} & \multicolumn{2}{c|}{} & \multicolumn{2}{c|}{} & \multicolumn{2}{c|}{}\tabularnewline
\hline 
\multicolumn{2}{|c||}{$\infty$} & \multicolumn{2}{c|}{$\sim\frac{d}{6}$} & \multicolumn{2}{c||}{$\frac{5}{6}=0.8333$} & \multicolumn{2}{c||}{} & \multicolumn{2}{c|}{} & \multicolumn{2}{c||}{$\frac{4}{3}=1.333$} & \multicolumn{2}{c|}{$\sim\frac{1}{2d}$} & \multicolumn{2}{c|}{$\frac{1}{2}$} & \multicolumn{2}{c|}{$\sim\frac{d}{6}(?)$}\tabularnewline
\hline 
\end{tabular}\hfill{} 
\end{table}

\section{Ground State Finite-Size Correction Exponents\label{sec:Ground-State-Finite-Size}}

Since simulations of statistical systems are bound to be conducted
at system sizes $N$ typically quite far from the thermodynamic limit
$N\to\infty$, it becomes essential to understand the corrections
entailed by such limitation. This is especially pertinent for spin
glasses beset with extra complexities such as NP-hardness at $T=0$
(or, similarly, the lack of equilibration at low but finite $T$)
or the additional burden of disorder averaging over many random samples
severely limiting $N$. Only rarely do such corrections decay fast
enough to reveal the thermodynamic behavior of an observable in a
simulation at a single, ``large-enough'' $N$. Instead, as we have
already seen for the stiffness in Sec. \ref{sec:Domain-Wall-Stiffness},
typically, sets of data need to be generated to glean the asymptotic
behavior for large sizes. To extrapolate for the value of an intensive
observable (like the ground-state energy density) it is then necessary
to have a handle on the nature of the finite-size corrections (FSC)
that have to be expected for the generated data \citep{Pal96,Palassini00,Young24}.
However, FSC are not only a technical necessity. Their behavior is
often closely related to other physical properties in the thermodynamic
limit via scaling relations \citep{Bouchaud03}. They can also be
instrumentalized, for instance, to assess the scaleability of optimization
heuristics \citep{Boettcher19,Boettcher23}.

\begin{figure}
\begin{minipage}[b][1\totalheight][s]{0.48\textwidth}%
\caption{\label{fig:FSC}{\small{}{} Plot of finite-size corrections to ground
state energies in bond-diluted lattice spin glasses (EA). For each
dimension $d$, ground state averages $e_{L}$ at increasing system
sizes $L$ were obtained at a convenient bond-density $p$. An asymptotic
fit (dashed lines) of that data according to Eq. (\ref{eq:FSC}) was
obtained. The resulting values for the finite-size corrections exponent
$\omega$ are listed in Tab. \ref{tab:Stiffness-exponents} and plotted
in Fig. \ref{fig:1overd}, suggesting that Eq. (\ref{eq:omega}) holds.}}

\vspace{0.9cm}
\end{minipage}\hfill{}%
\begin{minipage}[b][1\totalheight][s]{0.48\textwidth}%
\includegraphics[bb=10bp 20bp 740bp 550bp,clip,width=1\columnwidth,viewport=0bp 20bp 730bp 550bp]{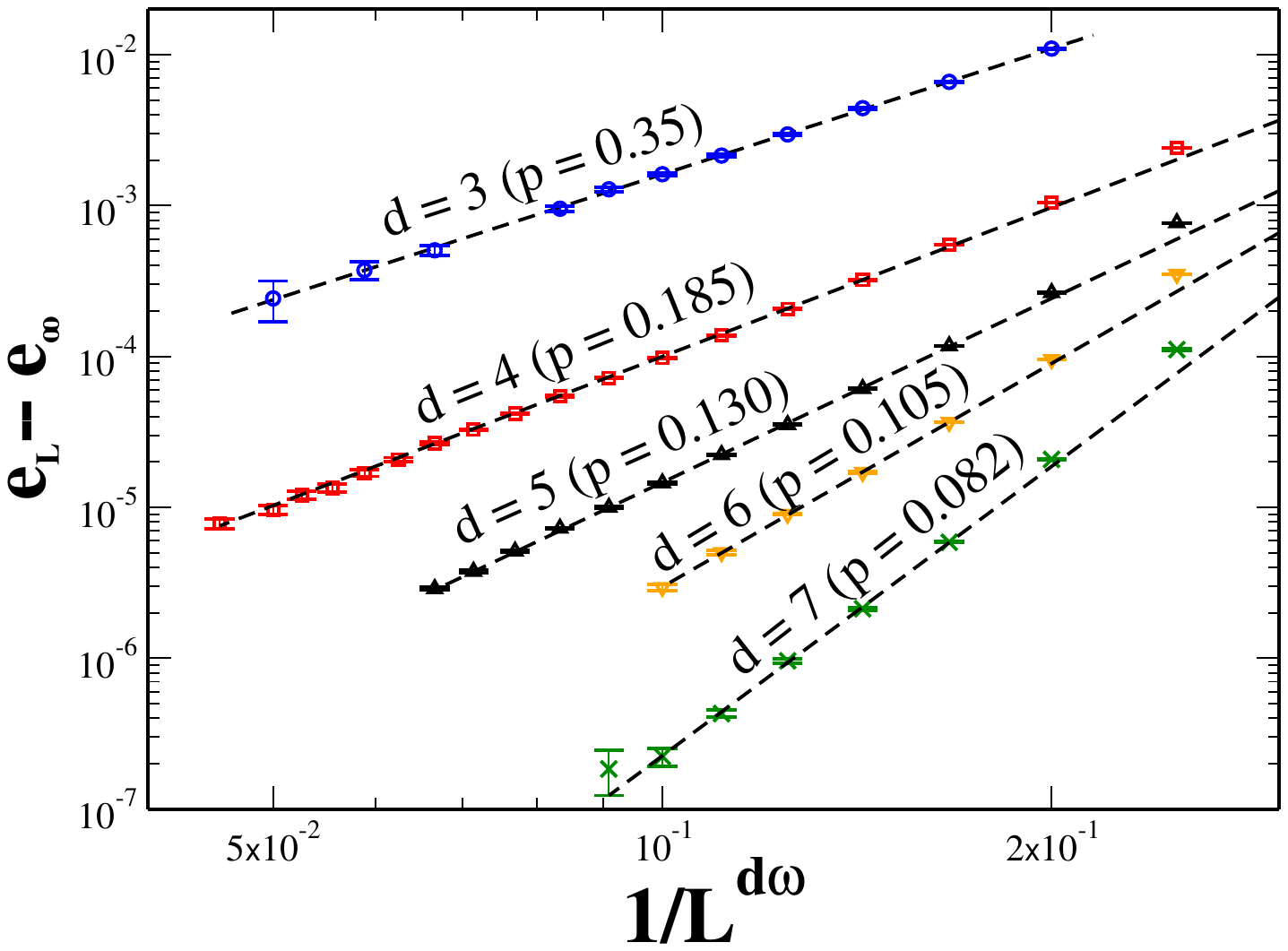}\hfill{}%
\end{minipage}\hfill{} 
\end{figure}

For the ground state energy densities in EA, Ref. \citep{Bouchaud03}
argued that such FSC should be due to locked-in domain walls of energy
$\sim L^{y}$, which would lead to the scaling correction for the
extensive energies of $E_{L}\sim e_{\infty}L^{d}+\Upsilon L^{y}$
for large $L$, defining $e_{\infty}$ as the $L\to\infty$ limit
of the average ground state energy density $e_{L}=\left\langle \frac{E_{L}}{L^{d}}\right\rangle $.
This is consistent with Eq. (\ref{eq:yscaling}), where we have purposefully
created such a domain wall, because the same system freed from that
domain wall (or locked into another one) would have $E_{L}^{\prime}\sim e_{\infty}L^{d}+\Upsilon^{\prime}L^{y}$,
thus, $\Delta E_{L}\sim\Delta\Upsilon L^{y}$. Dividing $E_{L}$ by
system size, we then get 
\begin{eqnarray}
e_{L} & \sim & e_{\infty}+\frac{A}{\left(L^{d}\right)^{\omega}},\qquad(L\to\infty).\label{eq:FSC}
\end{eqnarray}
where the FSC exponent is conjectured to be 
\begin{equation}
\omega=1-\frac{y}{d}.\label{eq:omega}
\end{equation}
Indeed, our direct evaluation of ground state energy densities at
some fixed bond density $p$ in dimensions $d=3,\ldots,7$ shown in
Fig. \ref{fig:FSC} are convincingly in agreement with this picture
for the dominant contributions to FSC. However, that leaves us with
somewhat of a conundrum when compared with mean-field simulations,
where FSC for the Sherrington-Kirkpatrick spin glass model (SK) \citep{EOSK,Boettcher10b,Aspelmeier07}
appear to yield $\omega\approx\frac{2}{3}$ for $d\to\infty$, which
is not close to $1-y/d\to\frac{5}{6}$ from RSB theory \citep{Parisi08}.

We conducted a corresponding ground state study at the edge of the
SG regime (see Fig. \ref{fig:PhaseDia}) by choosing the percolation
point $p=p_{c}$ exactly. Since the fractal percolation cluster cannot
sustain an ordered state, we find that the stiffness exponent defined
in Eq. (\ref{eq:yscaling}) is negative there, $y\vert_{p=p_{c}}=y_{P}<0$.
Numerical studies of ground states at $p_{c}$ (using Gaussian bonds
$J_{ij}$ in this case) is computationally quite efficient, since
the fractals embedded in the lattice reduce often completely or so
substantially that heuristics produce little systemic error. Large
lattice sizes $L$ can be achieved, limited only by rare large remainder
graphs or the lack of memory needed to build the original, unreduced
EA lattice. The values for $y_{P}$ thus obtained \citep{Boettcher07a}
are also listed in Tab. \ref{tab:Stiffness-exponents}. Although the
hypothesis for FSC from Eq. (\ref{eq:omega}), $\omega=1-y_{P}/d$,
leads to large values for $\omega$ when $y_{P}<0$ and it becomes
hard to test numerically, the corrections found are well consistent
with the hypothesis \citep{BoFa11}. In particular, it appears that
$1-y_{P}/d\to\frac{4}{3}$ for $d\geq d_{u}=6$, which would be consistent
with FSC in percolating random graphs \citep{Bollobas}. While this
provides an argument that Eq. (\ref{eq:omega}) should also hold in
the mean-field limit for EA in the spin glass phase, SK might be a
poor representation of that limit for EA. In EA, we first let $L\to\infty$
for fixed number of neighbors $2dp$ before $d\to\infty$ , while
in SK both system size and neighborhood diverge simultaneously. Unfortunately,
also sparse mean-field spin glasses on regular graphs (``Bethe lattices'')
appear to have FSC with $\omega=\frac{2}{3}$ \citep{Boettcher03a},
but those results might depend to some extend on structural details
of the mean-field networks \citep{Boettcher10b,Zdeborova10,Boettcher20}
and which structure most closely resembles a mean-field version of
EA at $d\to\infty$ remains unclear.

\begin{figure}
\begin{minipage}[b][1\totalheight][s]{0.48\textwidth}%
\caption{\label{fig:1overd}{\small{}{} Plot summarizing the data for the
exponents in Tab. \ref{tab:Stiffness-exponents}, here plotted as
a function of inverse dimension, $1/d$, to highlight the connection
with the mean field limit for $d\protect\geq d_{u}=6$ (left vertical
line). The bottom plot refers to the stiffness exponents $y$ in the
spin-glass regime (SG in Fig. \ref{fig:PhaseDia}) or $y_{P}$ at
$p_{c}$ and $T=0$, each presented as $1-y/d$. Included are also
the measured FSC exponents $\omega$, which appear to be consistent
with the conjecture in Eq. (\ref{eq:omega}). For stiffness, the $y$-data
is quite consistent with $1-y/d=5/6$ predicted for $d\protect\geq d_{u}$
\citep{Parisi08}, but not with the FSC $\omega_{{\rm SK}}=2/3$ found
for SK \citep{EOSK}. Fit of this data (solid line) yields a lower
critical dimension $d_{l}\approx\frac{5}{2}$, where $y=0$ (right
vertical line). At $p_{c}$, the $y_{P}$-data appears to approach
a value of $\omega=4/3$ expected for FSC of percolating random graphs.
In the top plot, $y_{P}$ is multiplied with the independent percolation
exponent $\nu$to form the thermal-percolative crossover exponent
$\phi$ that characterizes the behavior of the phase boundary near
$p_{c}$ in Eq. (\ref{eq:Tc}), see green arrow in Fig. \ref{fig:PhaseDia}.
It seems to show a minimum of $\phi\approx1$ at $d=d_{u}=6$.}}
\end{minipage}\hfill{}%
\begin{minipage}[b][1\totalheight][s]{0.48\textwidth}%
\includegraphics[clip,width=1.05\textwidth,viewport=10bp 40bp 730bp 250bp]{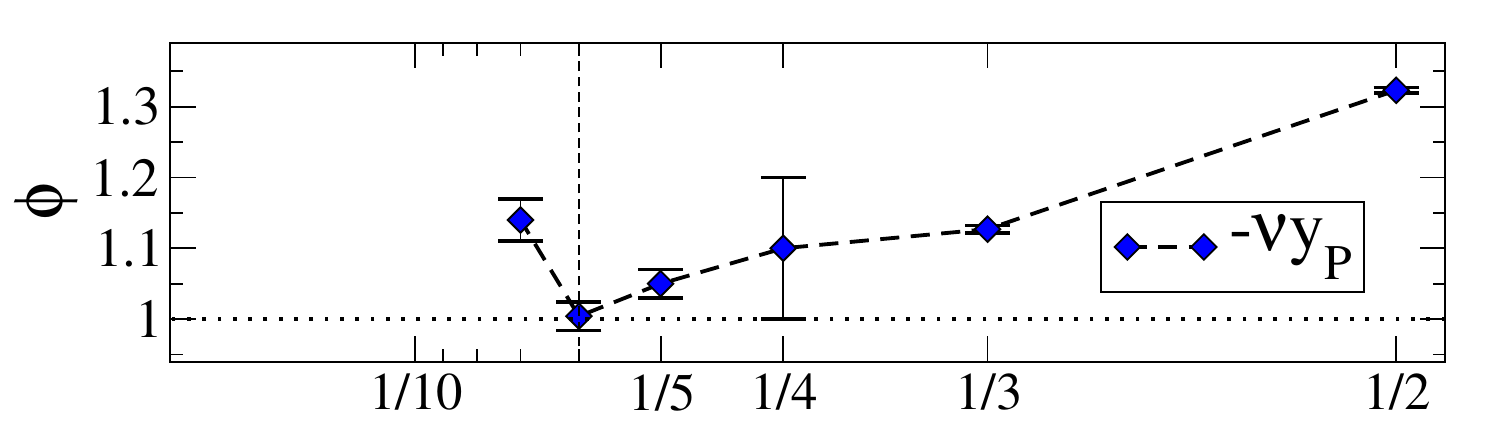}\hfill{}

\includegraphics[clip,width=1.05\textwidth,viewport=10bp 20bp 730bp 530bp]{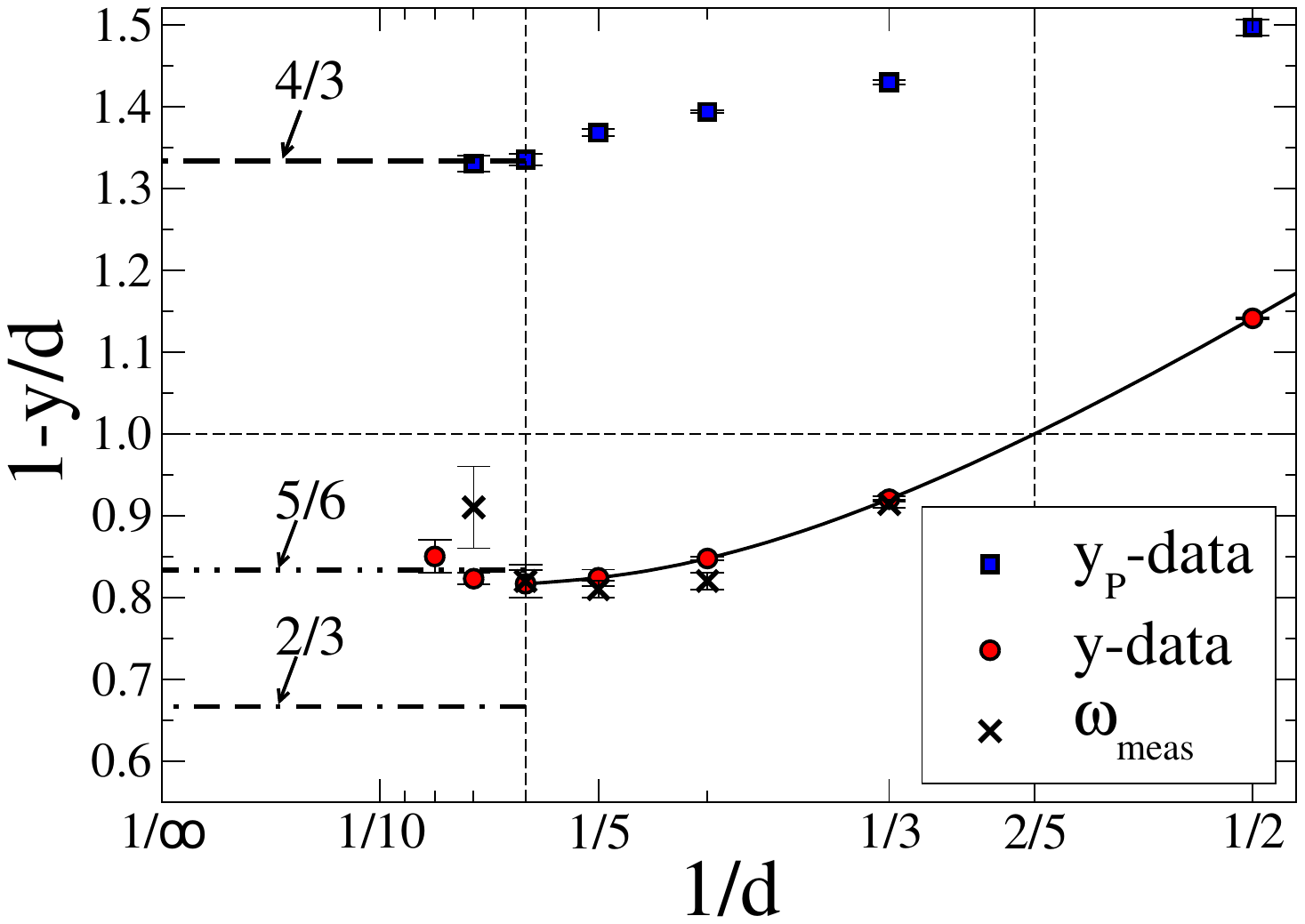}\hfill{}%
\end{minipage}\hfill{} 
\end{figure}

\section{Thermal-Percolative Crossover Exponents\label{sec:Thermal-Percolative-Crossover}}

Having determined the percolative stiffness exponents $y_{P}$ already
in the previous section, we can utilize it to make an interesting
-- and potentially experimentally testable -- prediction about the
behavior of the phase transition line in Fig. \ref{fig:PhaseDia}.
For diluted lattices at \emph{variable} bond density $p\to p_{c}$,
Eq. (\ref{eq:yscaling}) generalizes to \citep{Banavar87,Bray87b}
\begin{eqnarray}
\sigma(\Delta E)_{L,p}\sim{\cal Y}(p)L^{y}f\left(L/\xi(p)\right).\label{sigmaeq}
\end{eqnarray}
Here, we assume ${\cal Y}(p)\sim(p-p_{c})^{t}\sim\xi^{-t/\nu}$ for
the surface tension and $\xi(p)\sim(p-p_{c})^{-\nu}$ is the conventional
correlation length for percolation. The scaling function $f$ is defined
to be constant for $L\gg\xi(p)\gg1$, where percolation (and hence,
$\xi$) plays no role and we regain Eq.~(\ref{eq:yscaling}) for
$p>p_{c}$. For $\xi\gg L\gg1$, Eq.~(\ref{sigmaeq}) requires $f(x)\sim x^{\mu}$
for $x\to0$ to satisfy $\sigma\to0$ with some power of $L$, needed
to cancel the $\xi$-dependence at $p=p_{c}$. Thus, $\mu=-t/\nu$,
and we obtain $y_{P}=y+\mu=y-t/\nu$ to mark the $L$-dependence of
$\sigma$ at $p=p_{c}$, as before, which yields $t=\nu(y-y_{P})$.
Finally, at the cross-over $\xi\sim L$, where the range $L$ of the
excitations $\sigma(\Delta E)$ reaches the percolation length beyond
which spin glass order ensues, Eq.~(\ref{sigmaeq}) provides 
\begin{eqnarray}
\sigma(\Delta E)_{\xi(p),p}\sim\left(p-p_{c}\right)^{t}\xi(p)^{y}f(1)\sim\left(p-p_{c}\right)^{-\nu y_{P}}.\label{phieq}
\end{eqnarray}
Associating a temperature with the energy scale of this cross-over
by $\sigma(\Delta E)_{\xi(p),p}\sim T_{c}$ (since, for $T>T_{c}$,
thermal fluctuations destroy order at a length-scale $\ll\xi$), leads
to 
\begin{eqnarray}
T_{g}(p)\sim\left(p-p_{c}\right)^{\phi},\quad{\rm with}\quad\phi=-\nu y_{P},\label{eq:Tc}
\end{eqnarray}
defining \citep{Banavar87} the ``thermal-percolative cross-over
exponent'' $\phi$. All data for $d=2,\ldots,7$ are listed in Tab.~\ref{tab:Stiffness-exponents},
the results for $\phi$ are also shown in Fig. \ref{fig:1overd}.
It appears that $\phi$ declines with increasing $d$ for $d\leq d_{u}=6$,
has a minimum of $\phi=1$ at $d_{u}=6$, and rises as $\phi=d/6$
above $d_{u}$.

Of particular experimental interest is the result for $d=3$, $y_{P}=-1.289(6)$,
predicting $\phi=1.127(5)$ with $\nu=0.87436(46)$.\citep{Deng05}
This exponent provides a non-trivial, experimentally testable prediction
derived from scaling arguments of the equilibrium theory at low temperatures.
(Since bond and site percolation are typically in the same universality
class, it should make little difference whether an experiment varies
the site-concentration of atoms with dipolar spin or the bonds between
them.) Such tests are few as disordered materials by their very nature
fall out of equilibrium when entering the glassy state. The phase
boundary itself provides the perfect object for such a study: It can
be approached by theory from below and by experiments from above where
equilibration is possible. Ref.~\citep{Poon78} already provided
highly accurate results for the freezing temperature $T_{M}$ as a
function of dilution $x$ for a doped, crystalline glass, (La$_{1-x}$Gd$_{x})_{80}$Au$_{20}$,
proposing a linear dependence, $T_{M}\sim x$. The tabulated data
is equally well fitted by Eq.~(\ref{eq:Tc}) in that regime. Ref.~\citep{Beckman82}
determined a phase diagram for $({\rm Fe}_{x}{\rm Ni}_{1-x})_{75}{\rm P}_{16}{\rm B}_{6}{\rm Al}_{4}$,
an amorphous alloy, for a wide range of temperatures $T$ and site-concentrations
$x$ but did not discuss its near-linear behavior at low $x$. A similar
phase diagram for the insulator CdCr$_{2x}$In$_{2(1-x)}$S$_{4}$
can be found in Fig.~1.1a of Ref.~\citep{Vincent06}. New experiments
dedicated to the limit $x\searrow x_{c}$ should provide results of
sufficient accuracy to test our prediction for $\phi$.

\section{Conclusions\label{sec:Conclusions}}

We have summarized a collection of simulation data pertaining to the
lattice spin glass EA over a range of dimensions, providing a comprehensive
description of low-energy excitations from experimentally accessible
systems to the mean-field level, where exact results can be compared
with. Putting all those results side-by-side paints a self-consistent
picture of domain-wall excitations, their role in the stability of
the ordered glass state, and their role for finite-size corrections.
Extending to the very physical concept of bond-density made simulations
in high dimensions feasible, added accuracy, and opened up the spin-glass
phase diagram, which makes new observables experimentally accessible,
such as the thermal-percolative crossover exponent.

Going forward, the methods developed here could be extended to study,
say, ground state entropy and their overlaps \citep{MKpaper} or the
fractal nature of domain walls \citep{Wang17,Vedula24}. Our method
might inspire also new ways of using dilution as a gadget to make
simulations more efficient \citep{Jorg08}.  \bibliographystyle{apsrev}
\bibliography{../../../../../../../Boettcher}

\begin{thebibliography}{63}
\expandafter\ifx\csname natexlab\endcsname\relax\def\natexlab#1{#1}\fi
\expandafter\ifx\csname bibnamefont\endcsname\relax
  \def\bibnamefont#1{#1}\fi
\expandafter\ifx\csname bibfnamefont\endcsname\relax
  \def\bibfnamefont#1{#1}\fi
\expandafter\ifx\csname citenamefont\endcsname\relax
  \def\citenamefont#1{#1}\fi
\expandafter\ifx\csname url\endcsname\relax
  \def\url#1{\texttt{#1}}\fi
\expandafter\ifx\csname urlprefix\endcsname\relax\def\urlprefix{URL }\fi
\providecommand{\bibinfo}[2]{#2}
\providecommand{\eprint}[2][]{\url{#2}}

\bibitem[{\citenamefont{Wilson and Fisher}(1972)}]{Wilson72}
\bibinfo{author}{\bibfnamefont{K.~G.} \bibnamefont{Wilson}} \bibnamefont{and}
  \bibinfo{author}{\bibfnamefont{M.~E.} \bibnamefont{Fisher}},
  \bibinfo{journal}{Phys. Rev. Lett.} \textbf{\bibinfo{volume}{28}},
  \bibinfo{pages}{240} (\bibinfo{year}{1972}).

\bibitem[{\citenamefont{{t'Hooft} and Veltman}(1972)}]{tHooft72}
\bibinfo{author}{\bibfnamefont{G.}~\bibnamefont{{t'Hooft}}} \bibnamefont{and}
  \bibinfo{author}{\bibfnamefont{M.~J.~G.} \bibnamefont{Veltman}},
  \bibinfo{journal}{Nucl. Phys. B} \textbf{\bibinfo{volume}{44}},
  \bibinfo{pages}{189} (\bibinfo{year}{1972}).

\bibitem[{\citenamefont{Fischer and Hertz}(1991)}]{F+H}
\bibinfo{author}{\bibfnamefont{K.~H.} \bibnamefont{Fischer}} \bibnamefont{and}
  \bibinfo{author}{\bibfnamefont{J.~A.} \bibnamefont{Hertz}},
  \emph{\bibinfo{title}{Spin Glasses}}, Cambridge Studies in Magnetism
  (\bibinfo{publisher}{Cambridge University Press},
  \bibinfo{address}{Cambridge}, \bibinfo{year}{1991}).

\bibitem[{\citenamefont{{M\'ezard} et~al.}(1987)\citenamefont{{M\'ezard},
  Parisi, and Virasoro}}]{MPV}
\bibinfo{author}{\bibfnamefont{M.}~\bibnamefont{{M\'ezard}}},
  \bibinfo{author}{\bibfnamefont{G.}~\bibnamefont{Parisi}}, \bibnamefont{and}
  \bibinfo{author}{\bibfnamefont{M.~A.} \bibnamefont{Virasoro}},
  \emph{\bibinfo{title}{Spin glass theory and beyond}}
  (\bibinfo{publisher}{World Scientific}, \bibinfo{address}{Singapore},
  \bibinfo{year}{1987}).

\bibitem[{\citenamefont{Stein and Newman}(2013)}]{Stein13}
\bibinfo{author}{\bibfnamefont{D.~L.} \bibnamefont{Stein}} \bibnamefont{and}
  \bibinfo{author}{\bibfnamefont{C.~M.} \bibnamefont{Newman}},
  \emph{\bibinfo{title}{Spin Glasses and Complexity}}
  (\bibinfo{publisher}{Princeton University Press, Princeton},
  \bibinfo{year}{2013}).

\bibitem[{\citenamefont{Charbonneau et~al.}(2023)\citenamefont{Charbonneau,
  Marinari, and Mezard}}]{RSB40}
\bibinfo{editor}{\bibfnamefont{P.}~\bibnamefont{Charbonneau}},
  \bibinfo{editor}{\bibfnamefont{E.}~\bibnamefont{Marinari}}, \bibnamefont{and}
  \bibinfo{editor}{\bibfnamefont{M.}~\bibnamefont{Mezard}}, eds.,
  \emph{\bibinfo{title}{Spin glass theory and far beyond}}
  (\bibinfo{publisher}{World Scientific}, \bibinfo{address}{Singapore},
  \bibinfo{year}{2023}).

\bibitem[{\citenamefont{de~Dominicis et~al.}(1998)\citenamefont{de~Dominicis,
  Kondor, and Temes{\'a}ri}}]{dedominicis:98}
\bibinfo{author}{\bibfnamefont{C.}~\bibnamefont{de~Dominicis}},
  \bibinfo{author}{\bibfnamefont{I.}~\bibnamefont{Kondor}}, \bibnamefont{and}
  \bibinfo{author}{\bibfnamefont{T.}~\bibnamefont{Temes{\'a}ri}}, in
  \emph{\bibinfo{booktitle}{Spin Glasses and Random Fields}}, edited by
  \bibinfo{editor}{\bibfnamefont{A.}~\bibnamefont{Young}}
  (\bibinfo{publisher}{World Scientific}, \bibinfo{address}{Singapore},
  \bibinfo{year}{1998}).

\bibitem[{\citenamefont{Moore and Read}(2018)}]{Moore18}
\bibinfo{author}{\bibfnamefont{M.~A.} \bibnamefont{Moore}} \bibnamefont{and}
  \bibinfo{author}{\bibfnamefont{N.}~\bibnamefont{Read}},
  \bibinfo{journal}{Physical Review Letters} \textbf{\bibinfo{volume}{120}},
  \bibinfo{pages}{130602} (\bibinfo{year}{2018}).

\bibitem[{\citenamefont{Moore}(2021)}]{Moore21}
\bibinfo{author}{\bibfnamefont{M.~A.} \bibnamefont{Moore}},
  \bibinfo{journal}{Physical Review E} \textbf{\bibinfo{volume}{103}},
  \bibinfo{pages}{062111} (\bibinfo{year}{2021}).

\bibitem[{\citenamefont{Barahona}(1982)}]{Barahona82}
\bibinfo{author}{\bibfnamefont{F.}~\bibnamefont{Barahona}},
  \bibinfo{journal}{J. Phys. A: Math. Gen.} \textbf{\bibinfo{volume}{15}},
  \bibinfo{pages}{3241} (\bibinfo{year}{1982}).

\bibitem[{\citenamefont{Boettcher}(2004{\natexlab{a}})}]{Boettcher04b}
\bibinfo{author}{\bibfnamefont{S.}~\bibnamefont{Boettcher}},
  \bibinfo{journal}{The European Physical Journal B - Condensed Matter}
  \textbf{\bibinfo{volume}{38}}, \bibinfo{pages}{83}
  (\bibinfo{year}{2004}{\natexlab{a}}).

\bibitem[{\citenamefont{Boettcher}(2004{\natexlab{b}})}]{Boettcher04c}
\bibinfo{author}{\bibfnamefont{S.}~\bibnamefont{Boettcher}},
  \bibinfo{journal}{Europhys. Lett.} \textbf{\bibinfo{volume}{67}},
  \bibinfo{pages}{453} (\bibinfo{year}{2004}{\natexlab{b}}).

\bibitem[{\citenamefont{Boettcher}(2005{\natexlab{a}})}]{Boettcher05d}
\bibinfo{author}{\bibfnamefont{S.}~\bibnamefont{Boettcher}},
  \bibinfo{journal}{Physical Review Letters} \textbf{\bibinfo{volume}{95}},
  \bibinfo{pages}{197205} (\bibinfo{year}{2005}{\natexlab{a}}).

\bibitem[{\citenamefont{Boettcher and Marchetti}(2008)}]{Boettcher07a}
\bibinfo{author}{\bibfnamefont{S.}~\bibnamefont{Boettcher}} \bibnamefont{and}
  \bibinfo{author}{\bibfnamefont{E.}~\bibnamefont{Marchetti}},
  \bibinfo{journal}{Phys. Rev. B} \textbf{\bibinfo{volume}{77}},
  \bibinfo{pages}{100405(R)} (\bibinfo{year}{2008}).

\bibitem[{\citenamefont{Boettcher and Falkner}(2012)}]{BoFa11}
\bibinfo{author}{\bibfnamefont{S.}~\bibnamefont{Boettcher}} \bibnamefont{and}
  \bibinfo{author}{\bibfnamefont{S.}~\bibnamefont{Falkner}},
  \bibinfo{journal}{EPL (Europhysics Letters)} \textbf{\bibinfo{volume}{98}},
  \bibinfo{pages}{47005} (\bibinfo{year}{2012}).

\bibitem[{\citenamefont{Edwards and Anderson}(1975)}]{Edwards75}
\bibinfo{author}{\bibfnamefont{S.~F.} \bibnamefont{Edwards}} \bibnamefont{and}
  \bibinfo{author}{\bibfnamefont{P.~W.} \bibnamefont{Anderson}},
  \bibinfo{journal}{J. Phys. F} \textbf{\bibinfo{volume}{5}},
  \bibinfo{pages}{965} (\bibinfo{year}{1975}).

\bibitem[{\citenamefont{Boettcher and Davidheiser}(2008)}]{Boettcher08a}
\bibinfo{author}{\bibfnamefont{S.}~\bibnamefont{Boettcher}} \bibnamefont{and}
  \bibinfo{author}{\bibfnamefont{J.}~\bibnamefont{Davidheiser}},
  \bibinfo{journal}{Phys. Rev. B} \textbf{\bibinfo{volume}{77}},
  \bibinfo{pages}{214432} (\bibinfo{year}{2008}).

\bibitem[{\citenamefont{Boettcher and Percus}(2001)}]{Boettcher01a}
\bibinfo{author}{\bibfnamefont{S.}~\bibnamefont{Boettcher}} \bibnamefont{and}
  \bibinfo{author}{\bibfnamefont{A.~G.} \bibnamefont{Percus}},
  \bibinfo{journal}{Physical Review Letters} \textbf{\bibinfo{volume}{86}},
  \bibinfo{pages}{5211} (\bibinfo{year}{2001}).

\bibitem[{\citenamefont{Hartmann and Rieger}(2004)}]{Dagstuhl04}
\bibinfo{editor}{\bibfnamefont{A.}~\bibnamefont{Hartmann}} \bibnamefont{and}
  \bibinfo{editor}{\bibfnamefont{H.}~\bibnamefont{Rieger}}, eds.,
  \emph{\bibinfo{title}{New Optimization Algorithms in Physics}}
  (\bibinfo{publisher}{Wiley-VCH}, \bibinfo{address}{Berlin},
  \bibinfo{year}{2004}).

\bibitem[{\citenamefont{Southern and Young}(1977)}]{Southern77}
\bibinfo{author}{\bibfnamefont{B.~W.} \bibnamefont{Southern}} \bibnamefont{and}
  \bibinfo{author}{\bibfnamefont{A.~P.} \bibnamefont{Young}},
  \bibinfo{journal}{J.~Phys.~C: Solid State Phys.}
  \textbf{\bibinfo{volume}{10}}, \bibinfo{pages}{2179} (\bibinfo{year}{1977}).

\bibitem[{\citenamefont{McMillan}(1984)}]{McMillan84}
\bibinfo{author}{\bibfnamefont{W.~L.} \bibnamefont{McMillan}},
  \bibinfo{journal}{J.~Phys.~C: Solid State Phys.}
  \textbf{\bibinfo{volume}{17}}, \bibinfo{pages}{3179} (\bibinfo{year}{1984}).

\bibitem[{\citenamefont{Fisher and Huse}(1986)}]{Fisher86}
\bibinfo{author}{\bibfnamefont{D.~S.} \bibnamefont{Fisher}} \bibnamefont{and}
  \bibinfo{author}{\bibfnamefont{D.~A.} \bibnamefont{Huse}},
  \bibinfo{journal}{Phys. Rev. Lett.} \textbf{\bibinfo{volume}{56}},
  \bibinfo{pages}{1601} (\bibinfo{year}{1986}).

\bibitem[{\citenamefont{Bray and Moore}(1986)}]{bray:86}
\bibinfo{author}{\bibfnamefont{A.~J.} \bibnamefont{Bray}} \bibnamefont{and}
  \bibinfo{author}{\bibfnamefont{M.~A.} \bibnamefont{Moore}}, in
  \emph{\bibinfo{booktitle}{Heidelberg Colloquium on Glassy Dynamics and
  Optimization}}, edited by
  \bibinfo{editor}{\bibfnamefont{L.}~\bibnamefont{Van~Hemmen}}
  \bibnamefont{and}
  \bibinfo{editor}{\bibfnamefont{I.}~\bibnamefont{Morgenstern}}
  (\bibinfo{publisher}{Springer}, \bibinfo{address}{New York},
  \bibinfo{year}{1986}), p. \bibinfo{pages}{121}.

\bibitem[{\citenamefont{Krzakala and Martin}(2000)}]{Krzakala00}
\bibinfo{author}{\bibfnamefont{F.}~\bibnamefont{Krzakala}} \bibnamefont{and}
  \bibinfo{author}{\bibfnamefont{O.}~\bibnamefont{Martin}},
  \bibinfo{journal}{Phys. Rev. Lett.} \textbf{\bibinfo{volume}{85}},
  \bibinfo{pages}{3013} (\bibinfo{year}{2000}).

\bibitem[{\citenamefont{Palassini and Young}(2000)}]{Palassini00}
\bibinfo{author}{\bibfnamefont{M.}~\bibnamefont{Palassini}} \bibnamefont{and}
  \bibinfo{author}{\bibfnamefont{A.~P.} \bibnamefont{Young}},
  \bibinfo{journal}{Phys. Rev. Lett.} \textbf{\bibinfo{volume}{85}},
  \bibinfo{pages}{3017} (\bibinfo{year}{2000}).

\bibitem[{\citenamefont{Palassini et~al.}(2003)\citenamefont{Palassini, Liers,
  Juenger, and Young}}]{Palassini03}
\bibinfo{author}{\bibfnamefont{M.}~\bibnamefont{Palassini}},
  \bibinfo{author}{\bibfnamefont{F.}~\bibnamefont{Liers}},
  \bibinfo{author}{\bibfnamefont{M.}~\bibnamefont{Juenger}}, \bibnamefont{and}
  \bibinfo{author}{\bibfnamefont{A.~P.} \bibnamefont{Young}},
  \bibinfo{journal}{Physical Review B} \textbf{\bibinfo{volume}{68}},
  \bibinfo{pages}{064413} (\bibinfo{year}{2003}).

\bibitem[{\citenamefont{Bouchaud et~al.}(2003)\citenamefont{Bouchaud, Krzakala,
  and Martin}}]{Bouchaud03}
\bibinfo{author}{\bibfnamefont{J.-P.} \bibnamefont{Bouchaud}},
  \bibinfo{author}{\bibfnamefont{F.}~\bibnamefont{Krzakala}}, \bibnamefont{and}
  \bibinfo{author}{\bibfnamefont{O.~C.} \bibnamefont{Martin}},
  \bibinfo{journal}{Phys. Rev. B} \textbf{\bibinfo{volume}{68}},
  \bibinfo{pages}{224404} (\bibinfo{year}{2003}).

\bibitem[{\citenamefont{Aspelmeier et~al.}(2003)\citenamefont{Aspelmeier,
  Moore, and Young}}]{Aspelmeier03}
\bibinfo{author}{\bibfnamefont{T.}~\bibnamefont{Aspelmeier}},
  \bibinfo{author}{\bibfnamefont{M.~A.} \bibnamefont{Moore}}, \bibnamefont{and}
  \bibinfo{author}{\bibfnamefont{A.~P.} \bibnamefont{Young}},
  \bibinfo{journal}{Phys. Rev. Lett.} \textbf{\bibinfo{volume}{90}},
  \bibinfo{pages}{127202} (\bibinfo{year}{2003}).

\bibitem[{\citenamefont{Bray and Moore}(1984)}]{Bray84}
\bibinfo{author}{\bibfnamefont{A.~J.} \bibnamefont{Bray}} \bibnamefont{and}
  \bibinfo{author}{\bibfnamefont{M.~A.} \bibnamefont{Moore}},
  \bibinfo{journal}{J.~Phys.~C: Solid State Phys.}
  \textbf{\bibinfo{volume}{17}}, \bibinfo{pages}{L463} (\bibinfo{year}{1984}).

\bibitem[{\citenamefont{Franz et~al.}(1994)\citenamefont{Franz, Parisi, and
  Virasoro}}]{Franz94}
\bibinfo{author}{\bibfnamefont{S.}~\bibnamefont{Franz}},
  \bibinfo{author}{\bibfnamefont{G.}~\bibnamefont{Parisi}}, \bibnamefont{and}
  \bibinfo{author}{\bibfnamefont{M.~A.} \bibnamefont{Virasoro}},
  \bibinfo{journal}{J. Phys. I (France)} \textbf{\bibinfo{volume}{4}},
  \bibinfo{pages}{1657} (\bibinfo{year}{1994}).

\bibitem[{\citenamefont{Hartmann and Young}(2001)}]{Hartmann01}
\bibinfo{author}{\bibfnamefont{A.~K.} \bibnamefont{Hartmann}} \bibnamefont{and}
  \bibinfo{author}{\bibfnamefont{A.~P.} \bibnamefont{Young}},
  \bibinfo{journal}{Phys. Rev. B} \textbf{\bibinfo{volume}{64}},
  \bibinfo{pages}{180404(R)} (\bibinfo{year}{2001}).

\bibitem[{\citenamefont{Guchhait and Orbach}(2014)}]{Guchhait14}
\bibinfo{author}{\bibfnamefont{S.}~\bibnamefont{Guchhait}} \bibnamefont{and}
  \bibinfo{author}{\bibfnamefont{R.}~\bibnamefont{Orbach}},
  \bibinfo{journal}{Phys. Rev. Lett.} \textbf{\bibinfo{volume}{112}},
  \bibinfo{pages}{126401} (\bibinfo{year}{2014}).

\bibitem[{\citenamefont{Maiorano and Parisi}(2018)}]{Maiorano18}
\bibinfo{author}{\bibfnamefont{A.}~\bibnamefont{Maiorano}} \bibnamefont{and}
  \bibinfo{author}{\bibfnamefont{G.}~\bibnamefont{Parisi}},
  \bibinfo{journal}{Proceedings of the National Academy of Sciences}
  \textbf{\bibinfo{volume}{115}}, \bibinfo{pages}{5129} (\bibinfo{year}{2018}).

\bibitem[{\citenamefont{Hartmann}(2000)}]{Hartmann2000}
\bibinfo{author}{\bibfnamefont{A.~K.} \bibnamefont{Hartmann}},
  \bibinfo{journal}{Physical Review E} \textbf{\bibinfo{volume}{63}},
  \bibinfo{pages}{016106} (\bibinfo{year}{2000}).

\bibitem[{\citenamefont{Parisi and Rizzo}(2008)}]{Parisi08}
\bibinfo{author}{\bibfnamefont{G.}~\bibnamefont{Parisi}} \bibnamefont{and}
  \bibinfo{author}{\bibfnamefont{T.}~\bibnamefont{Rizzo}},
  \bibinfo{journal}{Phys. Rev. Lett.} \textbf{\bibinfo{volume}{101}},
  \bibinfo{pages}{117205} (\bibinfo{year}{2008}).

\bibitem[{\citenamefont{Lorenz and Ziff}(1998)}]{Lorenz98}
\bibinfo{author}{\bibfnamefont{C.~D.} \bibnamefont{Lorenz}} \bibnamefont{and}
  \bibinfo{author}{\bibfnamefont{R.~M.} \bibnamefont{Ziff}},
  \bibinfo{journal}{Phys. Rev. E} \textbf{\bibinfo{volume}{57}},
  \bibinfo{pages}{230} (\bibinfo{year}{1998}).

\bibitem[{\citenamefont{Grassberger}(2003)}]{Grassberger03}
\bibinfo{author}{\bibfnamefont{P.}~\bibnamefont{Grassberger}},
  \bibinfo{journal}{Phys. Rev. E} \textbf{\bibinfo{volume}{67}},
  \bibinfo{pages}{036101} (\bibinfo{year}{2003}).

\bibitem[{\citenamefont{Deng and {Bl\"ote}}(2005)}]{Deng05}
\bibinfo{author}{\bibfnamefont{Y.}~\bibnamefont{Deng}} \bibnamefont{and}
  \bibinfo{author}{\bibfnamefont{H.~W.~J.} \bibnamefont{{Bl\"ote}}},
  \bibinfo{journal}{Phys. Rev. E} \textbf{\bibinfo{volume}{72}},
  \bibinfo{pages}{016126} (\bibinfo{year}{2005}).

\bibitem[{\citenamefont{Hughes}(1996)}]{Hughes96}
\bibinfo{author}{\bibfnamefont{B.~D.} \bibnamefont{Hughes}},
  \emph{\bibinfo{title}{Random Walks and Random Environments}}
  (\bibinfo{publisher}{Oxford University Press}, \bibinfo{address}{Oxford},
  \bibinfo{year}{1996}).

\bibitem[{\citenamefont{Pal}(1996)}]{Pal96}
\bibinfo{author}{\bibfnamefont{K.~F.} \bibnamefont{Pal}},
  \bibinfo{journal}{Physica A} \textbf{\bibinfo{volume}{233}},
  \bibinfo{pages}{60} (\bibinfo{year}{1996}).

\bibitem[{\citenamefont{Young}(2024)}]{Young24}
\bibinfo{author}{\bibfnamefont{A.}~\bibnamefont{Young}},
  \emph{\bibinfo{title}{Finite-Size Scaling}} (\bibinfo{publisher}{World
  Scientific}, \bibinfo{year}{2024}), pp. \bibinfo{pages}{599--615}.

\bibitem[{\citenamefont{Boettcher}(2019)}]{Boettcher19}
\bibinfo{author}{\bibfnamefont{S.}~\bibnamefont{Boettcher}},
  \bibinfo{journal}{Physical Review Research} \textbf{\bibinfo{volume}{1}},
  \bibinfo{pages}{033142} (\bibinfo{year}{2019}).

\bibitem[{\citenamefont{Boettcher}(2023)}]{Boettcher23}
\bibinfo{author}{\bibfnamefont{S.}~\bibnamefont{Boettcher}},
  \bibinfo{journal}{Nature Communications} \textbf{\bibinfo{volume}{14}},
  \bibinfo{pages}{5658} (\bibinfo{year}{2023}).

\bibitem[{\citenamefont{Boettcher}(2005{\natexlab{b}})}]{EOSK}
\bibinfo{author}{\bibfnamefont{S.}~\bibnamefont{Boettcher}},
  \bibinfo{journal}{The European Physical Journal B}
  \textbf{\bibinfo{volume}{46}}, \bibinfo{pages}{501}
  (\bibinfo{year}{2005}{\natexlab{b}}).

\bibitem[{\citenamefont{Boettcher}(2010)}]{Boettcher10b}
\bibinfo{author}{\bibfnamefont{S.}~\bibnamefont{Boettcher}},
  \bibinfo{journal}{Journal of Statistical Mechanics: Theory and Experiment}
  \textbf{\bibinfo{volume}{2010}}, \bibinfo{pages}{P07002}
  (\bibinfo{year}{2010}).

\bibitem[{\citenamefont{Aspelmeier et~al.}(2008)\citenamefont{Aspelmeier,
  Billoire, Marinari, and Moore}}]{Aspelmeier07}
\bibinfo{author}{\bibfnamefont{T.}~\bibnamefont{Aspelmeier}},
  \bibinfo{author}{\bibfnamefont{A.}~\bibnamefont{Billoire}},
  \bibinfo{author}{\bibfnamefont{E.}~\bibnamefont{Marinari}}, \bibnamefont{and}
  \bibinfo{author}{\bibfnamefont{M.~A.} \bibnamefont{Moore}},
  \bibinfo{journal}{Journal of Physics A: Mathematical and Theoretical}
  \textbf{\bibinfo{volume}{41}}, \bibinfo{pages}{324008}
  (\bibinfo{year}{2008}).

\bibitem[{\citenamefont{Bollobas}(1985)}]{Bollobas}
\bibinfo{author}{\bibfnamefont{B.}~\bibnamefont{Bollobas}},
  \emph{\bibinfo{title}{Random Graphs}} (\bibinfo{publisher}{Academic Press},
  \bibinfo{address}{London}, \bibinfo{year}{1985}).

\bibitem[{\citenamefont{Boettcher}(2003{\natexlab{a}})}]{Boettcher03a}
\bibinfo{author}{\bibfnamefont{S.}~\bibnamefont{Boettcher}},
  \bibinfo{journal}{The European Physical Journal B - Condensed Matter}
  \textbf{\bibinfo{volume}{31}}, \bibinfo{pages}{29}
  (\bibinfo{year}{2003}{\natexlab{a}}).

\bibitem[{\citenamefont{Zdeborov{\'{a}} and Boettcher}(2010)}]{Zdeborova10}
\bibinfo{author}{\bibfnamefont{L.}~\bibnamefont{Zdeborov{\'{a}}}}
  \bibnamefont{and}
  \bibinfo{author}{\bibfnamefont{S.}~\bibnamefont{Boettcher}},
  \bibinfo{journal}{Journal of Statistical Mechanics: Theory and Experiment}
  \textbf{\bibinfo{volume}{2010}}, \bibinfo{pages}{P02020}
  (\bibinfo{year}{2010}).

\bibitem[{\citenamefont{Boettcher}(2020)}]{Boettcher20}
\bibinfo{author}{\bibfnamefont{S.}~\bibnamefont{Boettcher}},
  \bibinfo{journal}{Physical Review Letters} \textbf{\bibinfo{volume}{124}},
  \bibinfo{pages}{177202} (\bibinfo{year}{2020}).

\bibitem[{\citenamefont{Banavar et~al.}(1987)\citenamefont{Banavar, Bray, and
  Feng}}]{Banavar87}
\bibinfo{author}{\bibfnamefont{J.~R.} \bibnamefont{Banavar}},
  \bibinfo{author}{\bibfnamefont{A.~J.} \bibnamefont{Bray}}, \bibnamefont{and}
  \bibinfo{author}{\bibfnamefont{S.}~\bibnamefont{Feng}},
  \bibinfo{journal}{Phys. Rev. Lett.} \textbf{\bibinfo{volume}{58}},
  \bibinfo{pages}{1463} (\bibinfo{year}{1987}).

\bibitem[{\citenamefont{Bray and Feng}(1987)}]{Bray87b}
\bibinfo{author}{\bibfnamefont{A.~J.} \bibnamefont{Bray}} \bibnamefont{and}
  \bibinfo{author}{\bibfnamefont{S.}~\bibnamefont{Feng}},
  \bibinfo{journal}{Phys. Rev. B} \textbf{\bibinfo{volume}{36}},
  \bibinfo{pages}{8456} (\bibinfo{year}{1987}).

\bibitem[{\citenamefont{Poon and Durand}(1978)}]{Poon78}
\bibinfo{author}{\bibfnamefont{S.~J.} \bibnamefont{Poon}} \bibnamefont{and}
  \bibinfo{author}{\bibfnamefont{J.}~\bibnamefont{Durand}},
  \bibinfo{journal}{Phys. Rev. B} \textbf{\bibinfo{volume}{18}},
  \bibinfo{pages}{6253} (\bibinfo{year}{1978}).

\bibitem[{\citenamefont{Beckman et~al.}(1982)\citenamefont{Beckman, Figueroa,
  Gramm, Lundgren, Rao, and Chen}}]{Beckman82}
\bibinfo{author}{\bibfnamefont{O.}~\bibnamefont{Beckman}},
  \bibinfo{author}{\bibfnamefont{E.}~\bibnamefont{Figueroa}},
  \bibinfo{author}{\bibfnamefont{K.}~\bibnamefont{Gramm}},
  \bibinfo{author}{\bibfnamefont{L.}~\bibnamefont{Lundgren}},
  \bibinfo{author}{\bibfnamefont{K.~V.} \bibnamefont{Rao}}, \bibnamefont{and}
  \bibinfo{author}{\bibfnamefont{H.~S.} \bibnamefont{Chen}},
  \bibinfo{journal}{Phys. Scr.} \textbf{\bibinfo{volume}{25}},
  \bibinfo{pages}{726} (\bibinfo{year}{1982}).

\bibitem[{\citenamefont{Vincent}(2007)}]{Vincent06}
\bibinfo{author}{\bibfnamefont{E.}~\bibnamefont{Vincent}}, in
  \emph{\bibinfo{booktitle}{Ageing and the Glass Transition}}, edited by
  \bibinfo{editor}{\bibfnamefont{M.}~\bibnamefont{Henkel}},
  \bibinfo{editor}{\bibfnamefont{M.}~\bibnamefont{Pleimling}},
  \bibnamefont{and} \bibinfo{editor}{\bibfnamefont{R.}~\bibnamefont{Sanctuary}}
  (\bibinfo{publisher}{Springer}, \bibinfo{address}{Heidelberg},
  \bibinfo{year}{2007}), vol. \bibinfo{volume}{716} of
  \emph{\bibinfo{series}{Springer Lecture Notes in Physics}},
  \bibinfo{note}{condmat/063583}.

\bibitem[{\citenamefont{Boettcher}(2003{\natexlab{b}})}]{MKpaper}
\bibinfo{author}{\bibfnamefont{S.}~\bibnamefont{Boettcher}},
  \bibinfo{journal}{The European Physical Journal B - Condensed Matter}
  \textbf{\bibinfo{volume}{33}}, \bibinfo{pages}{439}
  (\bibinfo{year}{2003}{\natexlab{b}}).

\bibitem[{\citenamefont{Wang et~al.}(2017)\citenamefont{Wang, Moore, and
  Katzgraber}}]{Wang17}
\bibinfo{author}{\bibfnamefont{W.}~\bibnamefont{Wang}},
  \bibinfo{author}{\bibfnamefont{M.~A.} \bibnamefont{Moore}}, \bibnamefont{and}
  \bibinfo{author}{\bibfnamefont{H.~G.} \bibnamefont{Katzgraber}},
  \bibinfo{journal}{Physical Review Letters} \textbf{\bibinfo{volume}{119}}
  (\bibinfo{year}{2017}).

\bibitem[{\citenamefont{Vedula et~al.}(2024)\citenamefont{Vedula, Moore, and
  Sharma}}]{Vedula24}
\bibinfo{author}{\bibfnamefont{B.}~\bibnamefont{Vedula}},
  \bibinfo{author}{\bibfnamefont{M.~A.} \bibnamefont{Moore}}, \bibnamefont{and}
  \bibinfo{author}{\bibfnamefont{A.}~\bibnamefont{Sharma}},
  \emph{\bibinfo{title}{Evidence that the at transition disappears below six
  dimensions}} (\bibinfo{year}{2024}).

\bibitem[{\citenamefont{{J\"org} and Ricci-Tersenghi}(2008)}]{Jorg08}
\bibinfo{author}{\bibfnamefont{T.}~\bibnamefont{{J\"org}}} \bibnamefont{and}
  \bibinfo{author}{\bibfnamefont{F.}~\bibnamefont{Ricci-Tersenghi}},
  \bibinfo{journal}{Phys. Rev. Lett.} \textbf{\bibinfo{volume}{100}},
  \bibinfo{pages}{177203} (\bibinfo{year}{2008}).

\bibitem[{\citenamefont{Boettcher and Percus}(2000)}]{Boettcher00}
\bibinfo{author}{\bibfnamefont{S.}~\bibnamefont{Boettcher}} \bibnamefont{and}
  \bibinfo{author}{\bibfnamefont{A.}~\bibnamefont{Percus}},
  \bibinfo{journal}{Artificial Intelligence} \textbf{\bibinfo{volume}{119}},
  \bibinfo{pages}{275} (\bibinfo{year}{2000}).

\bibitem[{\citenamefont{Boettcher and Grigni}(2002)}]{eo_jam}
\bibinfo{author}{\bibfnamefont{S.}~\bibnamefont{Boettcher}} \bibnamefont{and}
  \bibinfo{author}{\bibfnamefont{M.}~\bibnamefont{Grigni}},
  \bibinfo{journal}{Journal of Physics A: Mathematical and General}
  \textbf{\bibinfo{volume}{35}}, \bibinfo{pages}{1109} (\bibinfo{year}{2002}).

\bibitem[{\citenamefont{Boettcher and Paczuski}(1996)}]{BoPa2}
\bibinfo{author}{\bibfnamefont{S.}~\bibnamefont{Boettcher}} \bibnamefont{and}
  \bibinfo{author}{\bibfnamefont{M.}~\bibnamefont{Paczuski}},
  \bibinfo{journal}{Physical Review E} \textbf{\bibinfo{volume}{54}},
  \bibinfo{pages}{1082} (\bibinfo{year}{1996}).

\bibitem[{\citenamefont{Boettcher}(2009)}]{LION3}
\bibinfo{author}{\bibfnamefont{S.}~\bibnamefont{Boettcher}}, in
  \emph{\bibinfo{booktitle}{Lecture Notes in Computer Science}}, edited by
  \bibinfo{editor}{\bibfnamefont{T.}~\bibnamefont{St{\"u}tzle}}
  (\bibinfo{publisher}{Springer Berlin Heidelberg}, \bibinfo{year}{2009}), vol.
  \bibinfo{volume}{5851}, pp. \bibinfo{pages}{1--14}.

\end{thebibliography}

\newpage{}

\appendix

\part*{Supplementary Material}

\section*{Appendix A: Bond-Diluted Spin Glasses\label{sec:Appendix-A:-Bond-Diluted}}

We have developed an exact algorithm that is capable of drastically
reducing the size of sparsely connected spin glass systems \citep{Boettcher04b},
leaving a much reduced remainder graph whose ground state can be approximated
with great accuracy using heuristics, such as EO. The algorithm and
its efficiency is studied in more detail elsewhere~\citep{Boettcher08a}.
An extension to simultaneously compute the entropy density and overlap
for sparse spin glass systems is discussed in Ref.~\citep{MKpaper}).
We focus here exclusively on the reduction rules for the energy at
$T=0$. These rules apply to general, purely quadratic Ising spin
glass Hamiltonians such as Eq. (\ref{eq:Heq}). The reductions effect
both spins and bonds, eliminating recursively all zero-, one-, two-,
and three-connected spins. These operations eliminate and add terms
to the expression in Eq.~(\ref{eq:Heq}), but leave it form-invariant.
Offsets in the energy along the way are accounted for by a variable
$H_{o}$, which is \textit{exact} for a ground state configuration.

\textit{Rule I:} An isolated spin can be ignored entirely.

\textit{Rule II:} A one-connected spin $i$ can be eliminated, since
its state can always be chosen in accordance with its neighboring
spin $j$ to satisfy the bond $J_{i,j}$. For its energetically most
favorable state we adjust $H_{o}:=H_{o}-|J_{i,j}|$ and eliminate
the term $-J_{i,j}\,\sigma_{i}\,\sigma_{j}$ from $H$.

\textit{Rule III:} A double bond, $J_{i,j}^{(1)}$ and $J_{i,j}^{(2)}$,
between two vertices $i$ and $j$ can be combined to a single bond
by setting $J_{i,j}=J_{i,j}^{(1)}+J_{i,j}^{(2)}$ or be eliminated
entirely, if the resulting bond vanishes. This operation is very useful
to lower the connectivity of $i$ and $j$ by one.

\textit{Rule IV:} Replacing a two-connected spin $i$ between some
spins $1$ and $2$, the graph obtains a new bond $J_{1,2}$, and
acquires an offset $H_{o}:=H_{o}-\Delta H$, by rewriting in Eq.~(\ref{eq:Heq})
\begin{eqnarray}
 & \sigma_{i}(J_{i,1}\sigma_{1}+J_{i,2}\sigma_{2})\leq\left|J_{i,1}\sigma_{1}+J_{i,2}\sigma_{2}\right|=J_{1,2}\sigma_{1}\sigma_{2}+\Delta H,\nonumber \\
 & J_{1,2}=\frac{1}{2}\left(\left|J_{i,1}+J_{i,2}\right|-\left|J_{i,1}-J_{i,2}\right|\right),\quad\Delta H=\frac{1}{2}\left(\left|J_{i,1}+J_{i,2}\right|+\left|J_{i,1}-J_{i,2}\right|\right).\label{eq:2con}
\end{eqnarray}

\textit{Rule V:} A three-connected spin $i$ can be reduced via a
``star-triangle'' relation, see Fig.~\ref{startri}: 
\begin{eqnarray}
 & J_{i,1}\,\sigma_{i}\,\sigma_{1}+J_{i,2}\,\sigma_{i}\,\sigma_{2}+J_{i,3}\,\sigma_{i}\,\sigma_{3}\leq
J_{1,2}\,\sigma_{1}\,\sigma_{2}+J_{1,3}\,\sigma_{1}\,\sigma_{3}+J_{2,3}\,\sigma_{2}\,\sigma_{3}+\Delta H,\nonumber \\
 & J_{1,2}=-A-B+C+D,\quad J_{1,3}=A-B+C-D,\quad J_{2,3}=-A+B+C-D,\nonumber \\
 & \Delta H=A+B+C+D,\qquad A=\frac{1}{4}\left|J_{i,1}-J_{i,2}+J_{i,3}\right|,\label{eq:3con}\\
 & B=\frac{1}{4}\left|J_{i,1}-J_{i,2}-J_{i,3}\right|,\quad C=\frac{1}{4}\left|J_{i,1}+J_{i,2}+J_{i,3}\right|,\quad D=\frac{1}{4}\left|J_{i,1}+J_{i,2}-J_{i,3}\right|.\nonumber 
\end{eqnarray}

\textit{Rule VI:} A spin $i$ (of any connectivity) for which the
absolute weight $|J_{i,j'}|$ of one bond to a spin $j'$ is larger
than the absolute sum of all its other bond-weights to neighboring
spins $j\not=j'$, i. e., it's a ``super-bond'' with 
\begin{eqnarray}
|J_{i,j'}| & > & \sum_{j\not=j'}|J_{i,j}|\label{eq:superbond}
\end{eqnarray}
such that bond $J_{i,j'}$ \textit{must} be satisfied in any ground
state. Then, spin $i$ is determined in the ground state by spin $j'$
and it as well as the bond $J_{i,j'}$ can be eliminated accordingly.
Here, we obtain $H_{0}:=H_{0}-|J_{i,j'}|$. All other bonds connected
to $i$ are simply reconnected with $j'$, but with reversed sign,
if $J_{i,j'}<0$.

This procedure is costly, and hence best applied after the other rules
are exhausted. But it can be highly effective for widely distributed
bonds, e.g., for Gaussian rather than bimodal ${\cal P}(J)$. In particular,
since neighboring spins may reduce in connectivity and become susceptible
to the previous rules again, an avalanche of further reductions may
ensue.

The bounds in Eqs.~(\ref{eq:2con}-\ref{eq:3con}) become \textit{exact}
when the remaining graph takes on its ground state. Reducing higher-connected
spins leads to (hyper-)bonds between multiple spins, unlike Eq.~(\ref{eq:Heq}),
and is not considered here.

After a recursive application of these rules, the original lattice
graph is either completely reduced (which is almost always the case
for bond densities $p<p_{c}$), in which case $H_{o}$ provides the
exact ground state energy already, or we are left with a highly reduced,
compact graph in which no spin has less than four connections. We
obtain the ground state of the reduced graph with EO, which together
with $H_{o}$ provides a very accurate approximation to the ground
state energy of the original diluted lattice instance.

\section*{Appendix B: Extremal Optimization Heuristic \label{sec:Appendix-B:-EO}}

For all of the results presented in this work, we have employed the
Extremal Optimization heuristic (EO), specifically $\tau-$EO \citep{Boettcher00,Boettcher01a,Dagstuhl04},
that performs a local search on an existing configuration $\vec{\sigma}$
by changing preferentially those $\sigma_{i}$ of ``bad'' fitness
$\lambda_{i}$. Here, the fitness for each spin $\sigma_{i}$ is defined
as the ratio of the sum of its satisfied weights over all its bond-weights,
$\lambda_{i}=\sum_{j}^{{\rm SAT}}\left|J_{i,j}\right|/\sum_{j}\left|J_{i,j}\right|$,
so that $0\leq\lambda_{i}\leq1$. EO \emph{ranks} variables by fitnesses,
$\lambda_{\Pi(1)}\leq\lambda_{\Pi(2)}\leq\ldots\leq\lambda_{\Pi(N)}$,
where the permutation $\Pi(k)=i$ is the index for the $k^{{\rm th}}$-ranked
$\sigma_{i}$, then randomly selects a rank with a \emph{scale-free}
probability $P_{k}\propto k^{-\tau}$. The selected variable $\sigma_{\Pi(k)}$
is updated \emph{unconditionally}, and it and all its neighbors reevaluate
and rerank their $\lambda_{i}$. For an easily determined choice\citep{Boettcher00,Boettcher01a}
of $\tau$ (for which theory \citep{eo_jam,Dagstuhl04} predicts as
optimal choice $\tau_{{\rm opt}}-1\sim\ln^{-1}N$), these unconditional
updates ensure a build-up of large, highly correlated fluctuations
through ``adaptive avalanches'' that \emph{learn} and \emph{memorize}
(as expressed within that ranking) what is favored in an optimal solutions
\citep{BoPa2,LION3}. It is the persistent bias against badly-adapted
variables that leads to frequent returns to near-optimal configurations.
As $\tau-$EO keeps fluctuating widely, it simply records the best-found
solution for the ground-state energy in passing, making it ideally
suited also to time-varying problems as well as annealing operations,
unencumbered by phase transitions~\citep{Boettcher19}.
\end{document}